\documentclass[a4paper,10pt]{article}
\usepackage{graphicx}
\usepackage{comment}
\usepackage{amssymb}
\usepackage{amsmath}
\usepackage{nicefrac}
\usepackage{graphicx}
\usepackage{dcolumn}
\usepackage{bm}
\usepackage{comment}
\usepackage{multirow}
\usepackage{cite}
\usepackage{url}
\usepackage[cyr]{aeguill}

\begin{document}

\huge

\begin{center}
Numerical modeling of isochoric heating experiments using the {\sc Troll} code in the warm dense matter regime
\end{center}

\vspace{0.3cm}

\large

\begin{center}
Sébastien Rassou$^{a,b,}$\footnote{sebastien.rassou@cea.fr}, Marie Bonneau$^{a}$, Christophe Rousseaux$^{a}$, Xavier Vaisseau$^{a}$, Witold Cayzac$^{a}$, Adrien Denoeud$^{a}$, Frédéric Perez$^{c}$, Tom Beaumont$^{a}$, Morris Demoulins$^{a}$ and Jean-Christophe Pain$^{a,b}$
\end{center}

\normalsize

\begin{center}
\it $^a$CEA, DAM, DIF, F-91297 Arpajon, France\\
\it $^b$Universit\'e Paris-Saclay, CEA, Laboratoire Mati\`ere en Conditions Extr\^emes,\\
\it 91680 Bruy\`eres-le-Ch\^atel, France\\
\it $^c$Laboratoire pour l'Utilisation des Lasers Intenses, Ecole Polytechnique, CNRS, Route de Saclay, 91128 Palaiseau, France

\end{center}

\vspace{0.3cm}

\begin{abstract}
Experiments of isochoric heating by protons of solid material were recently performed at LULI laser facilities. In these experiments, protons, produced from target normal sheath acceleration (TNSA) of Au foil with the PICO2000 laser, deposit their energy into an aluminum or copper foil initially at room temperature and solid density. The heated material reaches the warm dense matter regime with temperature in the rear face of the material between 1 and 5 eV. The temperature is inferred by streaked optical pyrometry and the proton beam is characterized by Thomson parabola. The high-energy protons produced by TNSA are modeled to deduce the initial proton distribution before the slowing down in the target. Hydrodynamic radiative simulations were next performed using the {\sc Troll} code in multidimensional geometry. In the {\sc Troll} code, the heating of protons is modeled with a Monte-Carlo transport module of charged particle and the calculation of the energy deposited by the protons in the matter is performed using stopping power formulas like {\sc Srim} functions. The results of simulations with the {\sc Troll} code are compared with the experimental results. An acceptable agreement between experiment and simulation is found for the temperature at the rear of the material using {\sc Sesame} equation of state and {\sc Srim} stopping power for protons in aluminum.
\end{abstract}

\section{Introduction}\label{sec1}

The study of warm dense matter (WDM) is particularly challenging due to the need to characterize its properties in temperature and density ranges that lie between those of condensed matter and plasma. This regime is critical for applications in astrophysics and inertial confinement fusion. However, under these conditions, conventional models either lose accuracy or become computationally prohibitive, necessitating experimental data to validate and refine them. In most cases, equation of state (EOS) models are extrapolated to this regime, introducing significant uncertainties.
Experimentally, the EOS of a material can be determined by rapidly heating it and performing precise measurements. For instance, in shock compression experiments, a shockwave—generated by radiatively heating a cavity—propagates through a sample material and a reference material. The pressure in the sample is then deduced using velocimetry measurements \cite{hicks2005, LMJ_CAYZAC2024101125}. However, such experiments only provide data along the Rankine-Hugoniot curve, which corresponds to conditions of high temperature and high pressure.
For moderate compressions or shockless conditions, alternative techniques such as electric current pulsing and Joule heating can be employed \cite{korobenko1999techniqueECP, JodarEPP_RSI}. These methods allow researchers to investigate material properties within the WDM regime. Additionally, isochoric heating—where the material is heated at constant volume—provides another avenue for studying EOS properties. In this context, isentropic expansion measurements, proposed by Foord \cite{foord2004determining}, have proven effective. For example, isochoric heating can occur in laser-matter interactions, where hot electrons confined by a self-generated magnetic field \cite{sentoku2007isochoricElectrons} rapidly heat the material on a picosecond timescale \cite{sawada2024IsoHeatingElec_XFEL}. To capture the resulting phenomena, ultrafast diagnostic tools such as X-ray free-electron lasers with sub-picosecond resolution are required.
Isochoric heating can also be achieved using proton beams\cite{snavely2007IsoAlu, mancic2010isochoric, HOARTY201250, feldman2017ProtonCopper, roycroft2020streaked, bhutwala2020development}, with the EOS measured through isentropic expansion, as described in Foord's formalism. This approach offers another valuable pathway for probing the complex properties of matter in the warm dense matter regime.
Isochoric heating means that the protons heat the material faster than it can expand. Typically, in the experiments performed at LULI (laboratoire pour l'utilisation des lasers intenses, Ecole Polytechnique, Palaiseau, France), the heating occurs in a few tens of picoseconds, which is faster than the hydrodynamic expansion (nanoseconds). Several challenges must be addressed before numerical modeling of isochoric heating by protons can be carried out. Proton production via target normal sheath acceleration (TNSA) and energy deposition resulting from proton stopping by electrons in bulk matter can only be accurately simulated using kinetic codes, such as particle-in-cell (PIC) codes \cite{carrie2010isochoric}, which operate on femtosecond time scales. Most recent models of isochoric heating by protons rely on 1D hydrodynamic codes, where proton energy deposition is typically pre-calculated \cite{mancic2010isochoric, bhutwala2020development, roycroft2020streaked}. In this work, some results of experiments performed at the LULI laser facility are compared with our simulation results with the {\sc Troll} code \cite{lefebvre2018development}, a radiation-hydrodynamics code. In the {\sc Troll} code, a charged particle transport module allows inline calculation of the energy deposition by a proton beam passing through matter. The results of {Man\v{c}i\'c \emph{et al.}} \cite{mancic2010isochoric} in the same type of experiments are used as reference for our modeling of the proton source. The choice of the stopping power of the proton is a key issue in modeling proton heating in matter. Depending on the material conditions and especially the degree of ionization of the material, various models are available \cite{deutsch2016ion, Barges2025Stop}. \\
The present work is divided into three parts. First, the experimental setup and results are briefly presented. In the second part, the modeling of the proton transport in matter with the {\sc Troll} code is explained. Finally, the comparison between simulation and experimental results is shown and discussed.

\section{Experimental setup and results}\label{sec2}

In this paper, the results obtained in the experimental campaign at the LULI laser facility in November 2021 are analyzed and compared with the {\sc Troll} simulation. In this section, a brief description of the experimental setup is presented. Some of the experimental results are shown. First, the proton distribution measured with a Thomson parabola is presented and then the radiative emission obtained by streaked optical pyrometry (SOP) is shown. The complete results of this experimental campaign will be presented in a separate paper.

\subsection{Experimental setup}

\begin{figure*}[!ht]
\centerline{\includegraphics[scale=0.60]{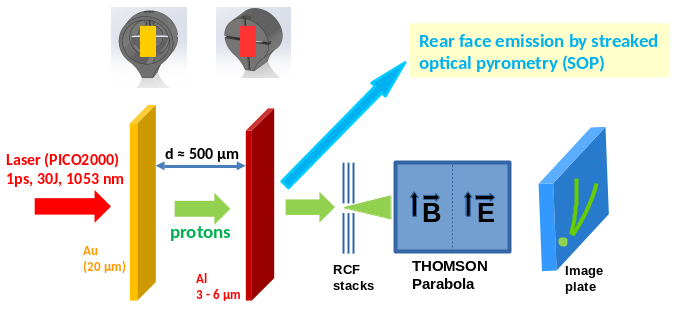}}
\caption{Scheme of the experimental setup of isochoric heating with protons experiments and the associated diagnostics for the proton beam (RCF and Thomson parabola) and radiative emission from the rear face (SOP). In the upper part of the picture, the plastic target holder is made with a 3D printer with the TNSA target on one side and the aluminum target on the other side. \label{Experimental_setup}}
\end{figure*}

The proton beam is generated through the TNSA mechanism when a picosecond laser is shot on a gold foil. In the experiments presented in this paper, the PICO2000 laser pulse is used with a duration of $1$ ps, an energy of around $30$ J and at $1053$ nm wavelength. The thickness of the gold foil is $20$ $\mu$m. The protons deposit energy in the sample placed at a distance $d$ of the gold foil. In the experiments, $d$=500 $\mu$m. We used samples of Al with thicknesses of 3 and 6 $\mu$m and a 300 $\mu$m width. The scheme of experimental setup is shown in Figure \ref{Experimental_setup}. Inside a vacuum chamber, the accelerated protons that pass through the target are diagnosed by the Thomson parabola and radiochromic films (RCF), while the radiative emission from the target is captured and imaged onto the slit of the SOP.

\begin{figure*}[!ht]
\centerline{\includegraphics[scale=0.65]{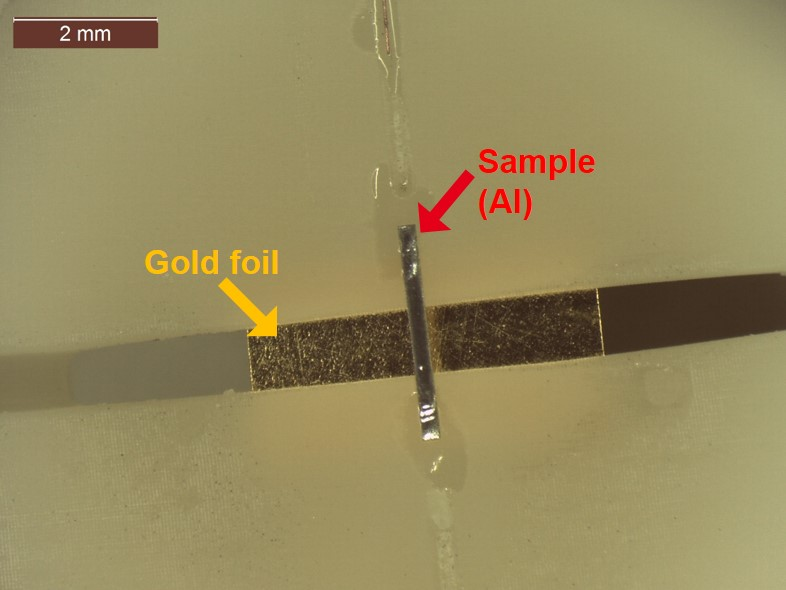}}
\caption{View of the central part of the target holder with the aluminum sample in the foreground and the gold foil in the background.} \label{Target}
\end{figure*}

The target for the proton beam generation and the sample are mounted on either side of a plastic target holder made with a 3D-printer (Figures \ref{Experimental_setup} and \ref{Target}) and the distance $d$ between the TNSA target and the sample is precisely measured.

The proton beam crossing the material passes through radiochromic films and the central part of the beam is collected by the Thomson parabola diagnostic using a device which combines magnets and metallic plates for the electric field.

\begin{figure*}[!ht]
\centerline{\includegraphics[scale=0.75]{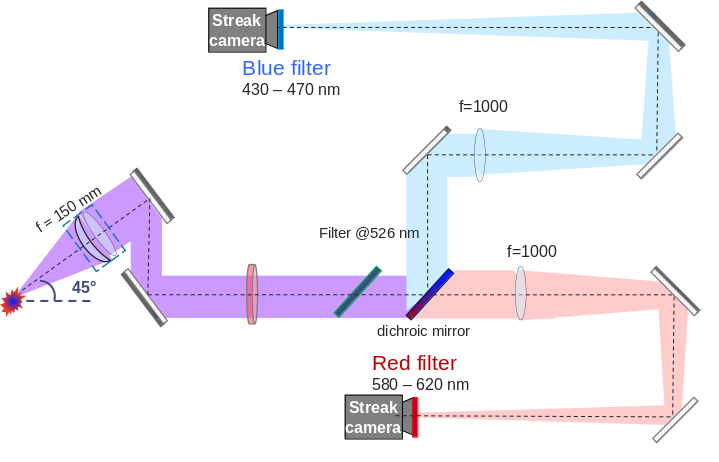}}
\caption{Scheme of the streaked optical pyrometry setup for collecting the radiative emission at the rear of the heated sample. Two streak cameras are used with appropriate filter to record the emission in two narrow regions of the spectrum. Expect for the first lens and the first mirror, the optical path lays in an horizontal plane.\label{SOP_setup}}
\end{figure*}

The time-dependent emissivity of the heated samples rear surface is recorded with a streaked optical pyrometry diagnostic. Two narrow wavelength regions of the spectrum are selected to cross-check the measurements. A scheme of the optical path and measurement principle is shown in Figure \ref{SOP_setup}.

Two streak cameras are used with appropriate filter to record the emission in two narrow regions of the spectrum. The ``red path'' corresponds to the wavelength 580-620 nm and the ``blue path'' to 430-470 nm.

In the following, two cases were analyzed with {\sc Troll} simulations and discussed. These are shots 19 and 40. The characteristics of the two laser shots are summarized in Table \ref{Table_expshot}.

\begin{table}[!ht]
\caption{Information on the analyzed experimental shots}
\centering
\begin{tabular*}{250pt}{@{\extracolsep\fill}lcc@{\extracolsep\fill}}
\textbf{Shot} & \textbf{19} & \textbf{40} \\\hline\hline
laser energy (J) & 35 & 35 \\
material & Al & Al \\
thickness ($\mu$m) & 3 & 6 \\
distance $d$ ($\mu$m) & 510 & 500 \\\hline\hline
\end{tabular*}
\label{Table_expshot}
\end{table}

More than 40 shots were made in the experimental campaign. For this analysis, we selected only two of them, for which we have a complete set of measurements, including the energy distribution at the center of the proton beam and the radiative emission of the rear face, such that they can be reliably analyzed and simulated.

\subsection{Proton energy distribution measurements}

In this work, the proton energy distributions are measured with Thomson parabola. It combines a magnetic field and an electric field region to deflect protons and other charged particles. Each type of charged particle is separated from the others because of its charge-to-mass ratio and measured on the imaging plate (Figure \ref{Experimental_setup}). The resulting trace for each ion type is a parabola depending on the value of the magnetic and electric fields, the distances and the incident energy of the particles. \\

Imaging plate of type BAS-TR 2025 from FUJI Photo Film Corporation were used in the Thomson parabola. They were scanned with a FUJI Film BAS-1800 II scanner to determine the photostimulated luminescence (PSL) count on the imaging plate. The conversion from PSL to proton number is performed using the process described by Rabhi \cite{rabhi2017calibration} for protons in the energy range $[0.1, 200]$ MeV.\\
For each shot, the protons pass through a stack of radiochromic films with an aperture in the center, allowing the central part of the proton beam to be measured using Thomson parabola diagnostic, as shown in Figure \ref{Experimental_setup}. In this work, the radiochromic films are not analyzed to determine the transverse profile. However, this provides information on whether the central part of the beam is collected in the Thomson parabola.\\

\begin{figure*}[!ht]
\centerline{\includegraphics[scale=0.45]{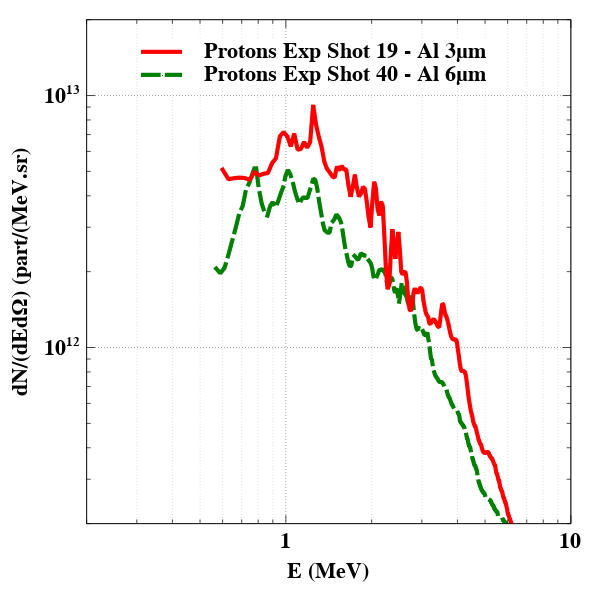}}
\caption{Proton energy distribution $\frac{dN}{dEd\Omega}$(part.MeV$^{-1}$.sr$^{-1}$) measurements by Thomson parabola for shots 19 (in solid red line) and 40 (in dotted green line). \label{ProtonMeasures}}
\end{figure*}

The proton energy distributions for shots 19 and 40 are shown in Figure \ref{ProtonMeasures}.

The two proton distributions are close for the high energy part of the distribution. For the low energy part, a significant difference is observed due to the material thickness and the larger deposition of the proton energy in 6 $\mu$m aluminum foil than in 3 $\mu$m. The lowest mesurable proton energy is 0,6 MeV and is due to the finite length of the Imaging Plate.

\subsection{Temperature measurements}

\begin{figure*}[!ht]
\centerline{\includegraphics[scale=0.5]{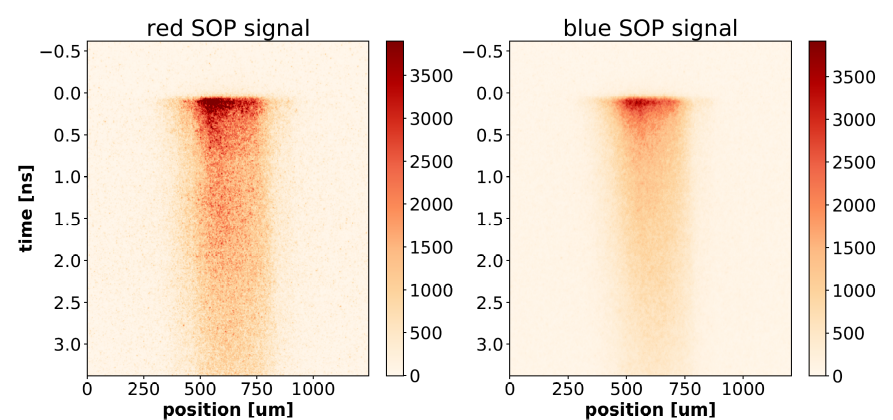}}
\caption{Streak camera image obtained for the red path (left) and the blue path (right) for shot 19 with streaked optical pyrometry as a function of the position along one axis and the time. The colormap unit is arbitrary. \label{Experimental_SOP_19}}
\end{figure*}

In Figure \ref{Experimental_SOP_19}, the result obtained with the SOP diagnostic for the shot 19 is shown. The emission is recorded as a function of time and in one direction in space and for both paths. 

\begin{figure*}[t]
\centerline{\includegraphics[scale=0.6]{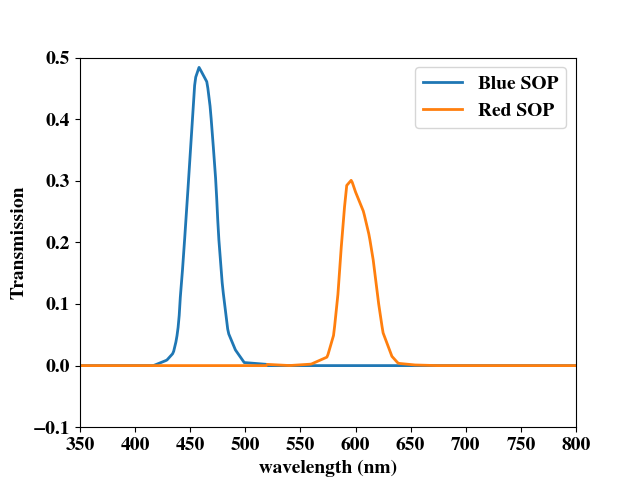}}
\caption{Spectral transmission of the two SOP optical paths, between the first lens and the streak cameras (blue line for the blue SOP path and orange for the red SOP one) as a function of the wavelength (nm). \label{Transmission_SOP}}
\end{figure*}

In order to calibrate the photometry of the SOP paths, we used a light concentrator source  [18] (LCS), whose emission is stable over the streak camera sweep duration. 
Immediately after the experimental campaign, we placed the LCS at the target position and recorded the spectral optical power at the fiber output of the LCS and in front of the streak cameras, as well as the corresponding streak camera signal. This calibration method was later on refined and described by Nourry-Martin et al. \cite{nourry2023absolute}. For the experimental campaign that we analyze in this paper, the calibration suffered from two major limitations. First, while the Ce:YAG LCS (described in Reference\cite{nourry2023absolute}) is perfectly suited for the red SOP path, it does not cover the spectral region of the blue SOP path. Second, the imaging of the 3 mm diameter fiber output face through the optical paths shows significant vignetting. The streak camera images of the red path provide information about the power spatial distribution along the camera horizontal axis. The signal is rather flat over a 100 pixel-wide central zone, corresponding to about 150 $\mu$m at the target surface,  where we assume that no vignetting takes place. On both sides of this zone, the signal decreases linearly. Unfortunately we do not have a direct measurement of the vignetting effect along the camera vertical axis. Therefore, we must make assumptions to derive the optical power entering the camera slit from the total optical power measured in front of the camera. Taking into account the horizontal vignetting profile, we then calculate the incident power in the 100 pixel central zone. From the average signal in this zone for different sweep durations, we estimate the single pixel mean sensitivity of the red path streak camera to be 0.17 nW/lsb (-43\% ; +88\%) for this central area and for the 5 ns sweep duration used for shots 19 and 40. The lower error bar value corresponds to a situation without vertical vignetting while the upper one corresponds to a vertical vignetting similar to the horizontal one. Furthermore, it should be noted that according to the calibration described in reference \cite{nourry2023absolute} (performed on a different setup with the same streak camera and the same slit height), one would expect the sensitivity to be much higher (of the order of 2.7 nW/lsb), a discrepancy that we cannot explain. Based on our estimate of the sensitivity of the streak camera, we obtain the calibration of the red SOP path from the measured magnification of the red optical path (8.6 horizontally, 7.2 vertically) and from its spectral transmission, whose theoretical value is plotted in Figure \ref{Transmission_SOP} and was confirmed by the spectrophotometer measurements.\\

\begin{figure*}[t]
\centerline{\includegraphics[scale=0.45]{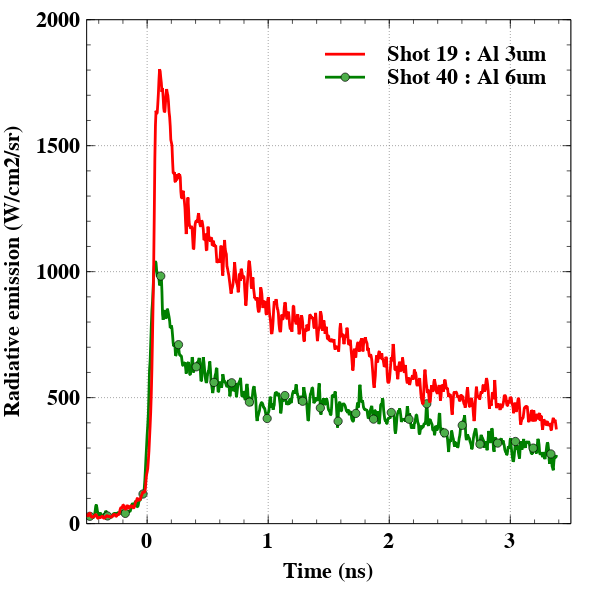} \includegraphics[scale=0.45]{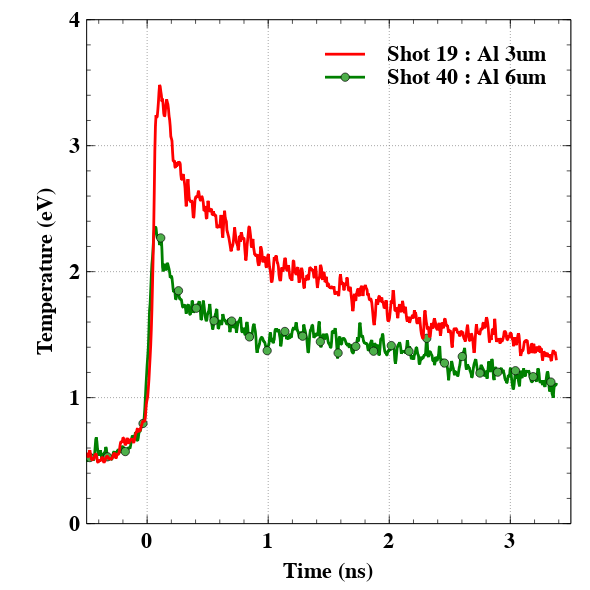}}
\caption{On the left, the radiative emission measurements by streaked optical pyrometry (``red'' SOP path) and on the right the temperature estimated assuming a black body radiation.The red line and the green line with circle correspond to shots 19 and 40 respectively. \label{Emission_Temp_shot19_40_Exp}}
\end{figure*}

Finally, the radiative emission normal to the target surface is calculated from the experimental data (shown in Figure \ref{Experimental_SOP_19} for shot 19), taking into account for the 0.028 sr collection solid angle of the first lens and for the 45° angle to the target surface (see Figure \ref{SOP_setup}). The time-dependent radiative emission is spatially averaged over typically 100 $\mu$m around the emission center. The results for shots 19 and 40 are shown in Figure \ref{Emission_Temp_shot19_40_Exp} (left). Combining the uncertainty of the vertical vignetting effect and the uncertainties on magnification, collection angle and direction, we estimate the error bars on the emission to be of the order of [-50\% , +100\%].\\

From the streaked optical pyrometer emission measurements at the rear face of the heated target, the temperature is estimated with the black body emission formula : 

\begin{equation}
B\ \left( \lambda , T \right) = 2\frac{\epsilon hc^{2}}{\lambda^5} \left( \frac{1}{\exp\left(\frac{hc}{k_bT\lambda}\right)-1}\right)
\label{eqBCN}
\end{equation}
with $B$ the radiative emission in $\left(\textrm{W.cm}^{-2}\textrm{.sr}^{-1}\textrm{.nm}^{-1}\right)$, $h$ the Planck constant, $c$ the speed of light, $\lambda$ the photon wavelength, $k_b$ the Boltzmann constant and $T$ the temperature.\\

When converting the SOP measurement to a temperature, the emissivity $\epsilon$ is set equal to $1$ based on calculations from Celliers and Ng \cite{celliers1993optical, roycroft2020streaked} which show that $\epsilon \approx 1$ for conditions similar to this experiment. We can notice that the formula is not valid for low temperature (< 1 eV) or very low radiative emission.\\

\begin{figure*}[t]
\centerline{\includegraphics[scale=0.6]{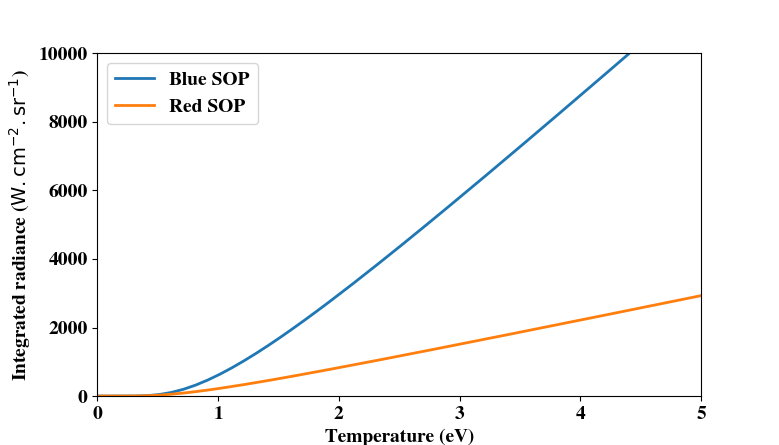}}
\caption{Integrated radiance $\left(\textrm{W.cm}^{-2}\textrm{.sr}^{-1}\right)$ for the two SOP paths and with associated transmission as a function of the temperature (eV). \label{RadEmissionSOP_functionT}}
\end{figure*}

The wavelength-integrated radiance scaled by the transmission function Figure \ref{Transmission_SOP}) is calculated as a function of the temperature \cite{roycroft2020streaked} using the Eq.(\ref{eqBCN}) and plotted in Figure \ref{RadEmissionSOP_functionT}.

Using these data, the temperature of the rear face of the material is estimated and plotted in Figure \ref{Emission_Temp_shot19_40_Exp} (right).\\

The modeling of the experiments is presented in the next section. Comparisons between the experimental measurements and the simulations are performed to test our measurement processing approach.

\section{Modeling proton transport in matter with {\sc Troll}}\label{sec3}

The {\sc Troll} code is used to model the experiments of the isochoric heating by protons performed at LULI. {\sc Troll} is a radiation-hydrodynamics code developed at CEA DAM to perform simulation of Hohlraum heating by laser for instance \cite{lefebvre2018development} or model implosion experiments in the indirect drive inertial confinement scheme \cite{liberatore2023first}. The {\sc Troll} code is initially designed to model matter in extreme conditions and at high energy densities. Having a realistic modeling of warm dense matter in the code is important and challenging, and isochoric heating experiments will provide a good test to benchmark the code in these conditions of temperature and density. 

\subsection{The {\sc Troll} code and the proton transport}

\begin{figure*}[!ht]
\centerline{\includegraphics[scale=0.66]{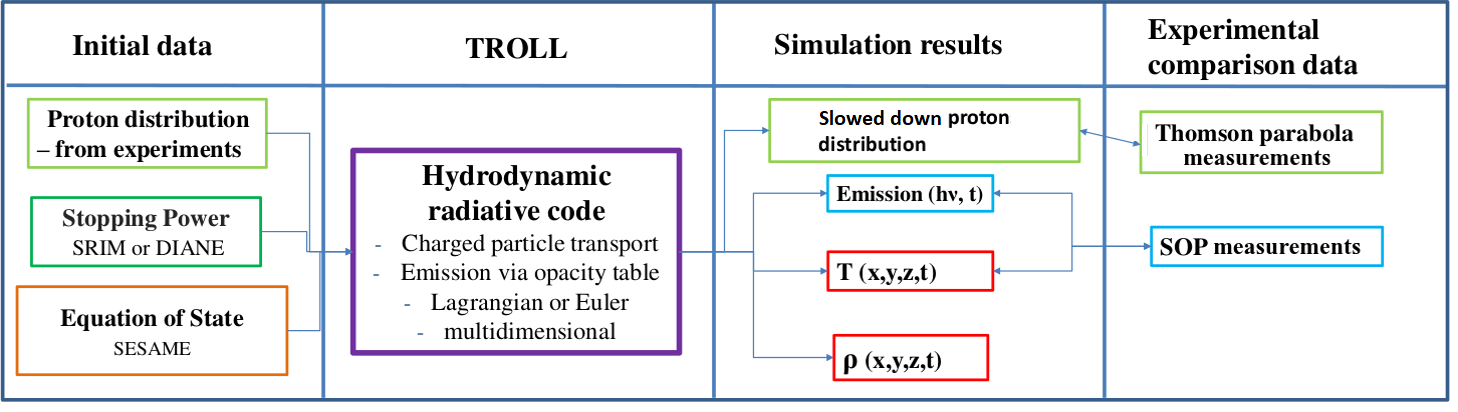}}
\caption{The different steps of modeling isochoric heating by proton experiments with {\sc Troll}. $\rho$ is matter density (usually in $\textrm{g.cm}^{-3}$). \label{Troll_fonctions}}
\end{figure*}

In {\sc Troll}, many features are available to model the matter in extreme conditions of temperature and pressure. It combines many modules of physics as well as equation of state and opacity databases. The transport of charged particles in matter is modeled with a dedicated module with several stopping power models available. The scheme in Figure \ref{Troll_fonctions} shows how the code proceeds during a simulation, and the various modules and data involved for modeling the experiments of isochoric heating by protons.

First, the initial data are chosen as entry in the code. The {\sc Sesame} equation of state \cite{mchardy2018introductionSESAME} is selected in our simulations. For the proton input data, the incident proton distribution is extracted from the measurements of the Thomson parabola and two stopping powers models are considered for comparison, the first based on the well-known {\sc Srim} \cite{ziegler2010srim} functions and the second being the model included in the {\sc Diane} code \cite{caillaud2014diane}. The radiation-transfer module and the associated opacity table can also be used to model the emission of the material during the isochoric heating of the material, but they do not have a significant effect on the temperature evolution in the material, where the major part of the heating is due to the energy deposition by the protons \cite{mancic2010isochoric}.

\begin{figure*}[t]
\centerline{\includegraphics[scale=0.35]{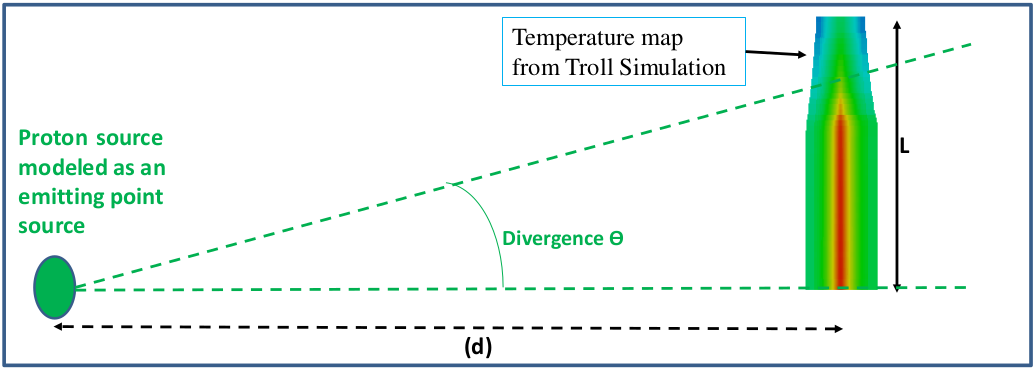}}
\caption{Schematic of the protons transport modeling in {\sc Troll}. A temperature map is also shown as the result of the heating of the proton beam in aluminum sample with a beam divergence $\theta$. \label{proton_transport}}
\end{figure*}

In {\sc Troll}, proton source is modeled as a point source emitting protons with an energy and an angular distribution specified by the user. The protons transport from the emitting point is carried out by a Monte-Carlo simulation as described in reference of the above mentioned {\sc Diane} code \cite{caillaud2014diane}. In the picture of Figure \ref{proton_transport}, the temperature map obtained with {\sc Troll} simulation of an aluminum foil (3 $\mu$m) heated (at $t$=110 ps after the first protons reach the target) with a proton source shows the impact of the beam divergence in the heating by protons. Therefore, the heating of the sample can be treated in a multi-dimensional geometry. In our case, the 2D cylindrical geometry allows us to correctly model the heating of the sample by the proton beam and to compare the simulation results with experimental measurements.\\

At the end of the simulation, the transmitted proton energy distribution is compared with the Thomson parabola measurements, and the temperature obtained by the simulation is compared with the experimental temperature estimated by the SOP measurements. By post-processing the simulation with the radiation calculation module included in {\sc Diane} \cite{liberatore2023first, caillaud2014diane}, the sample emission and the SOP emission measurements are compared.

\subsection{Proton slowing down in the matter}

The average rate of energy deposition per unit length of the proton in the material is defined as the stopping power and various models are available for the proton stopping power in matter. The stopping power $S = -\frac{dE}{dx}$ is divided into three parts corresponding to the contribution of three types of particles.
\begin{equation}\label{threecom}
    S = -\frac{dE}{dx}= -\frac{dE}{dx}\Bigl|_{\bf I} -\frac{dE}{dx}\Bigl|_{\bf B} -\frac{dE}{dx}\Bigl|_{\bf F} \approx -\frac{dE}{dx}\Bigl|_{\bf B} -\frac{dE}{dx}\Bigl|_{\bf F},
\end{equation}\\
the subscripts $I$, $B$ and $F$ correspond respectively to the contributions of ions, bound electrons and free electrons. As written in Eq. (\ref{threecom}), the ionic part of the stopping power is generally negligible compared to bound and free electron part.\\

We now study the effect of the stopping power modeling. As a first step, two limit models of stopping power are used for comparison : first, the {\sc Srim} \cite{ziegler2010srim} function which concerns the cold non-ionized matter and {\sc Diane} model \cite{caillaud2014diane} for the fully ionized matter. This paper focuses on the modeling of the isochoric heating of protons using the code {\sc Troll} with the complete proton transport from the source and the heating by the protons for comparison with the experiment of the resulting temperature. More precise stopping powers including partial target ionization could be used in further investigations \cite{Barges2025Stop}.\\

\begin{figure*}[!ht]
\centerline{\includegraphics[scale=0.45]{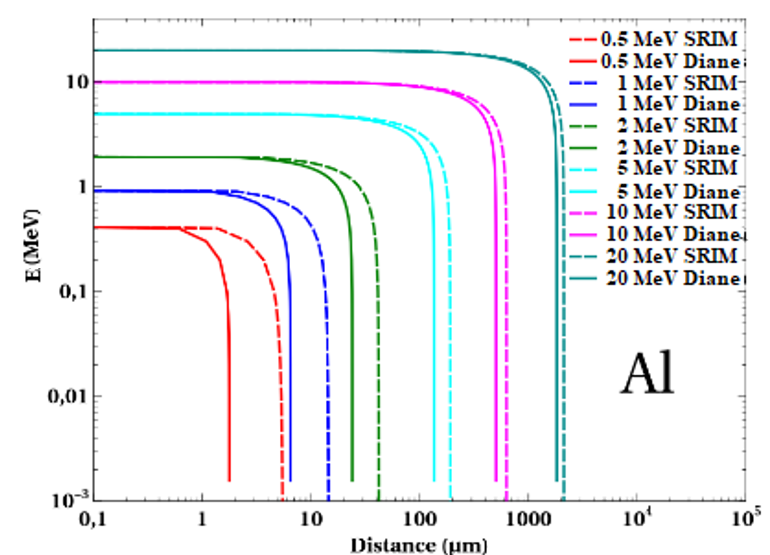}}
\caption{Protons energy (MeV) as a function of the distance crossed in aluminum foil for {\sc Srim} (dotted line) and {\sc Diane} (solid line) stopping power and for several initial proton energies.\label{SRIMvsDiane}}
\end{figure*}

In Figure \ref{SRIMvsDiane}, the proton energy curves in aluminum are shown for both stopping power models. The models give different results for protons with initial energy below 5 MeV. In our experiments, the foil is a few micrometer thick and the choice of the stopping power model is important to simulate properly the proton energy deposition and the heating. The resulting heating also depends on the accuracy of the input proton distribution at energies below 1 MeV.

\subsection{Modeling the input protons data from experiments}

To model isochoric heating by protons experiments with the {\sc Troll} code, the initial proton distribution has to be determined as input for the code. Some assumptions have been made to model the initial proton distribution reaching the sample, using only Thomson parabola measurements. The angular distribution of the protons is assumed to be homogeneous and the maximum angular divergence is assumed to be identical to {Man\v{c}i\'c \emph{et al.}}'s experimental fit of the proton beam \cite{mancic2010isochoric}, that have been obtained using TNSA with a gold foil.\\

\begin{figure*}[!ht]
\centerline{\includegraphics[scale=0.45]{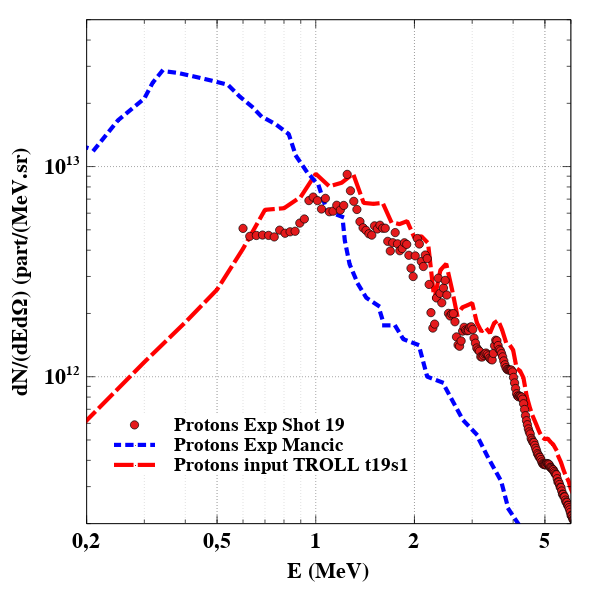} \includegraphics[scale=0.45]{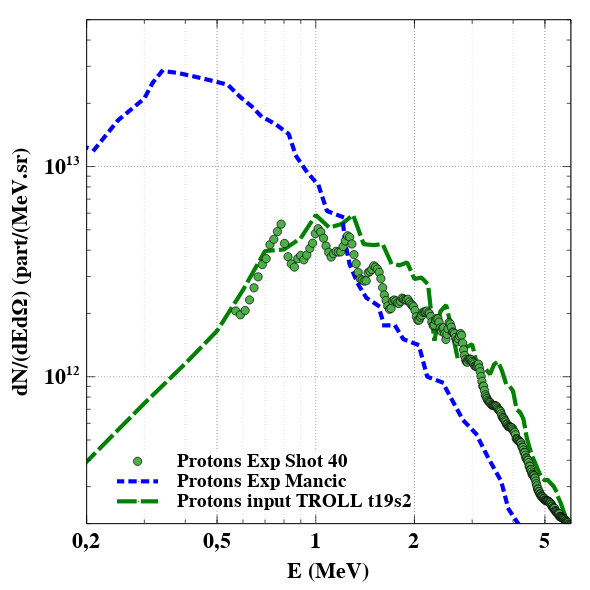}}
\caption{Proton distributions (in part/MeV/sr) used as input for the code to model the shot 19 (left, red dotted line) and the shot 40 (right, green dotted line) compared to the experimental results of the shots 19 (red circle) and 40 (green circle) obtained with Thomson parabola. The {Man\v{c}i\'c \emph{et al.}} experimental proton distribution (blue dotted line) is also plotted for comparison. \label{Protons_exp19_t19s_Mancic}}
\end{figure*}

There are two limitations with our Thomson parabola measurements of the proton distribution. First, the minimum energy of our measurement is 0.6 MeV, which is insufficient for significant proton energy deposition in micrometric aluminum foil as most of the energy deposition is due to protons with energies < 0.5 MeV. Second, the measurements are performed on transmitted protons and not incident protons.\\
Therefore, the proton distribution for low energies is extrapolated by following the shape of the distribution measured by {Man\v{c}i\'c \emph{et al.}} in these experiments \cite{mancic2010isochoric}. Then, using the {\sc Srim} or {\sc Diane} stopping powers, the incident proton distribution is estimated by deconvoluting the slowing down in the experimental conditions of the material and its thickness. In addition to the already described corrections applied on the measurements, the incident proton distribution might be tuned to match the temperature measurements. For shot 19, the proton number is multiplied by a  factor f19=1.25 and the resulting energy distribution is called ``t19s1''. For simplicity, the input data denoted ``t19s2'' for the simulation of shot 40 is also derived from the experimental proton distribution of shot 19, but the total number of protons is adjusted  to fit the high energy (> 4 MeV) part of the experimental proton distribution of shot 40 by applying a scaling factor f40=0.75. The input proton distribution are displayed in Figure \ref{Protons_exp19_t19s_Mancic} for shot 19 and for shot 40 and compared to the experimental ones. 

Under the conditions of shots 19 and 40, the estimated incident proton distributions are very close to the measured transmitted proton distributions.

\section{Simulation results and discussion}\label{sec4}

The experiments of isochoric heating by protons are simulated in 2D cylindrical dimension using {\sc Troll} with a Lagrangian mesh. In the 2D cylindrical simulation, the size of the aluminum sample is 150 $\mu$m in the transverse direction (y) and 3 $\mu$m or 6 $\mu$m in the proton beam propagation direction (x) for shot 19 and shot 40 respectively . The mesh contains 300$\times$100 cells for shot 19 and 300$\times$200 cells for shot 40 with the same intial cell size ($\Delta$y=0.5 $\mu$m and $\Delta$x=0.03 $\mu$m).  As described in the previous section, the proton beam is defined as an input parameter derived from the experimental results of the Thomson parabola diagnostic. The emitting proton point source coincides with the Au foil surface at an experimentally measured distance d from the target (Table \ref{Table_expshot}).\\

\begin{figure*}[!ht]
\centerline{\includegraphics[scale=0.45]{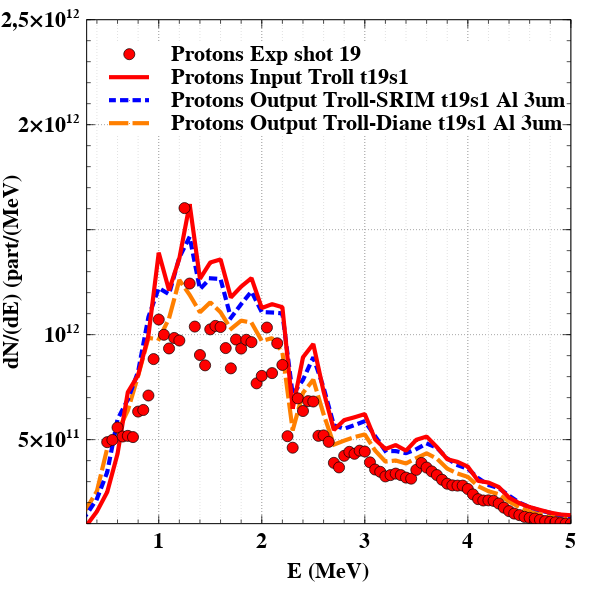} \includegraphics[scale=0.45]{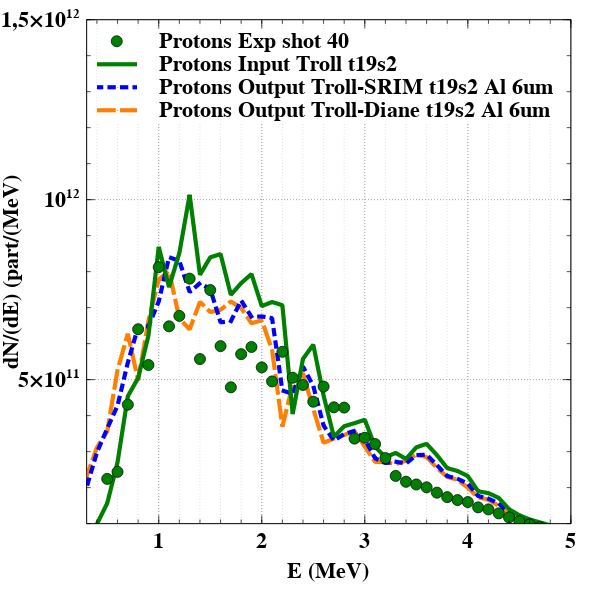}}
\caption{Proton distribution (in part/MeV) obtained by simulation for the two considered proton stopping power models and compared to experimental measurements and to the input distribution. The results for shot 19 are on the left and on the right for shot 40. \label{Protons_t19_t40_SRIMvsDiane}}
\end{figure*}

To check the consistency between simulation and experiment, the output proton distribution obtained with the {\sc Troll} simulation is then compared to the experimental distribution (Figure \ref{Protons_t19_t40_SRIMvsDiane}). The agreement is poor for low proton energies, but the experimental uncertainties on the Thomson parabola measurements are too large to conclude. As expected, the initial proton distribution is not significantly altered by the energy deposition in the sample material.\\

\begin{figure*}[!ht]
\centerline{\includegraphics[scale=0.45]{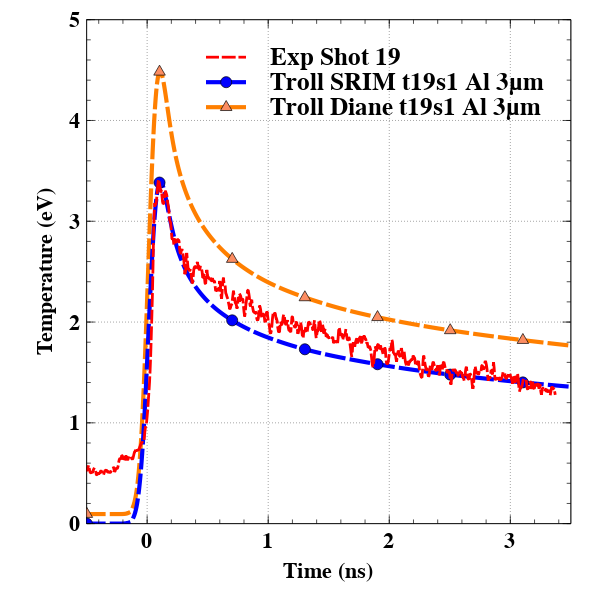} \includegraphics[scale=0.45]{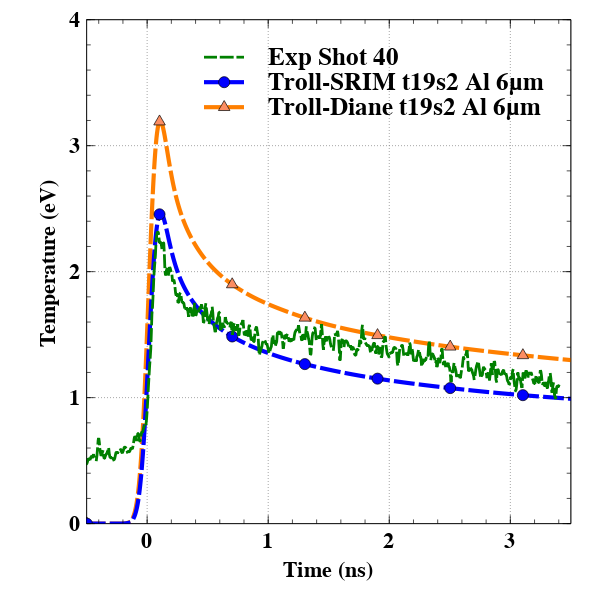}}
\caption{Temperature obtained by simulation for the two considered proton stopping power models and compared to the temperature estimated with a black body assumption from the experimental results for shot 19 (left) and for shot 40 (right). \label{Temperature_t19_t40_SRIMvsDiane}
}
\end{figure*}

In Figure \ref{Temperature_t19_t40_SRIMvsDiane}, we compare the temperatures obtained from the simulations with the Troll code with the estimated experimental temperatures for shots 19 and 40. The simulated temperature is averaged over the rear surface of the sample which corresponds to last cells in the mesh (along the transverse direction (y) and initially at x=3 $\mu$m). Therefore, we can compare it with the temperature estimated with our SOP setup and the black body assumption. The agreement between simulation and experiment is good for the two shots when the {\sc Srim} stopping is considered.

\begin{figure*}[!ht]
\centerline{\includegraphics[scale=0.45]{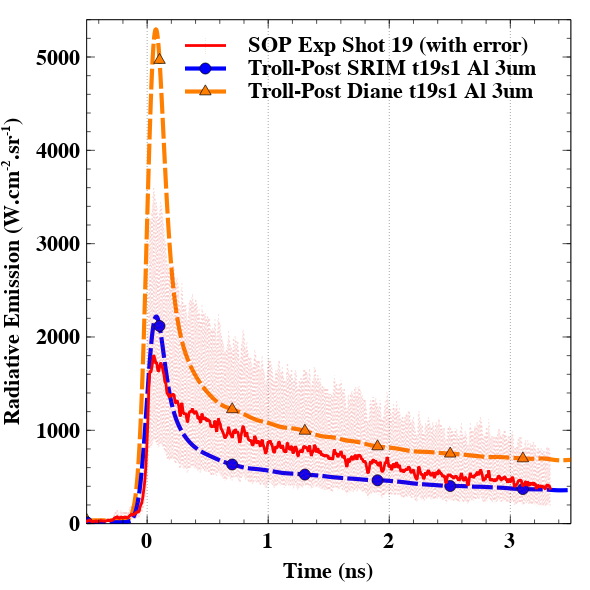} \includegraphics[scale=0.45]{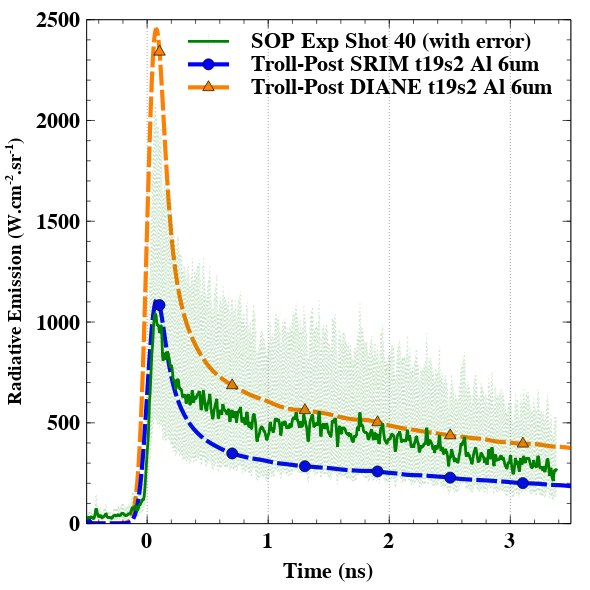}}
\caption{Radiative emission obtained by simulation for the two considered proton stopping power models and compared to experimental measurements. The results for shot 19 are on the left and on the right for shot 40. The error bars are estimated to be [$-50\textrm{\%}$, $+100\textrm{\%}$] on the measurements. \label{Emission_t19_t40_SRIMvsDiane}}
\end{figure*}

The simulations are then post-processed to determine the radiative emission and to compare with the SOP measurements (Figure \ref{Emission_t19_t40_SRIMvsDiane}). The post-processing consists in solving the radiative-transfer equation using opacity and equation-of-state tables, in order to determine the radiative emission at the rear face. Thanks to the 2D cylindrical geometry, post-processing is performed in conditions close to the experimental measurements by the SOP.\\

We can see that the radiative emission simulated with the {\sc Srim} stopping power is very close to the measurements. However, the emission calculated using the {\sc Diane} stopping power lies in the error bars as well, due to their rather large magnitude (of the order of -50\%/+100\%). The experimental radiative emission ratio between shots 19 and 40 is similar to the simulated ones. However, the temporal evolution is not consistent with the measurements. The relevance of the opacity tables in the wavelength range and at the considered temperature is questionable.

\section{Conclusions}\label{sec5}

In this work, the numerical modeling of the isochoric heating of a sample by a proton beam is presented. For this purpose, the hydrodynamic radiative code, {\sc Troll} is used to model recent experiments performed at LULI. The {\sc Troll} simulations are compared with experimental results. We first derive the initial proton energy distribution from experimental data measured with a Thomson parabola by applying a simple model to deconvolute the slowing down in the material. Due to missing data at low energy, the proton distribution is corrected according to the experimental data of {Man\v{c}i\'c \emph{et al.}} and the beam divergence is obtained from a function proposed by {Man\v{c}i\'c \emph{et al.}}. The next step is to compare the temperatures between simulations and experiments. In the experiments, the temperature is deduced with a black body assumption of material radiative emission at the rear face.
The good agreement shows that the latter assumption is reliable in the case of aluminum radiative emission in the range of 1 to 5 eV and in the expansion plasma regime. Two stopping powers models are also compared. Simulations with the cold matter {\sc Srim} stopping power are in better agreement with the experiments than simulations with the {\sc Diane} stopping power for fully ionized plasma. It is difficult to determine the best stopping power model in our case due to the uncertainties in our experimental Thomson parabola measurements and difficulties in estimating the radiative emission by simulation in the considered temperature range and the emission wavelength of the diagnostics. The {\sc Troll} code is shown to model rather correctly the isochoric heating by protons experiments, where the transport of the protons and the deposition of their energy give results close to the experimental measurements by comparing the temperatures.

\bibliographystyle{unsrt}
\bibliography{wileyNJD-AMA}

\end{document}